%% file: EM_3D_Processing_v4_submitted.tex
\documentclass[10pt, final, twocolumn, twoside, romanappendices]{IEEEtran}

\IEEEoverridecommandlockouts

\usepackage{graphicx,cite,acronym,amsmath,amssymb,amsfonts,color,subfigure,floatflt,stfloats,bm,textgreek,multirow}
\usepackage{epsfig,epstopdf,psfrag}
\usepackage[ruled,vlined]{algorithm2e}
\SetKwInput{KwInput}{Input}                
\SetKwInput{KwOutput}{Output}  

\usepackage[table]{xcolor}
\usepackage{bm}

\usepackage[cmintegrals]{newtxmath}
\usepackage{graphics,hhline,array,cite,bbm}

\usepackage{latexsym}
\usepackage{theorem}
\usepackage{times}
\usepackage{makeidx}
\DeclareGraphicsExtensions{.png,.jpg,.eps}

\input{acronyms}

\include{def_EM_SP}

\begin{document}
\title{Dynamic Scattering Arrays for Simultaneous Electromagnetic Processing and Radiation in Holographic MIMO Systems}

\author{
\IEEEauthorblockN{Davide~Dardari,~\IEEEmembership{Senior~Member,~IEEE}}
\IEEEcompsocitemizethanks{\IEEEcompsocthanksitem 
 D.~Dardari is with the 
   Dipartimento di Ingegneria dell'Energia Elettrica e dell'Informazione ``Guglielmo Marconi"  (DEI), WiLAB-CNIT, 
   University of Bologna, Cesena Campus, 
   Cesena (FC), Italy, (e-mail: davide.dardari@unibo.it). 
    }
}

\maketitle

\begin{abstract}

%

To meet the stringent requirements of next-generation wireless networks, \ac{MIMO} technology is expected to become massive and pervasive. Unfortunately, this could pose scalability issues in terms of complexity, power consumption, cost, and processing latency. Therefore, novel technologies and design approaches, such as the recently introduced holographic \ac{MIMO} paradigm, must be investigated to make future networks sustainable.

In this context, we propose the concept of a \ac{DSA} as a versatile 3D structure capable of performing joint wave-based computing and radiation by moving the processing from the digital domain to the \ac{EM} domain. We provide a general analytical framework for modeling \acp{DSA}, introduce specific design algorithms, and apply them to various use cases. The examples presented in the numerical results demonstrate the potential of \acp{DSA} to further reduce complexity and the number of \ac{RF} chains in holographic \ac{MIMO} systems while achieving enhanced \ac{EM} wave processing and radiation flexibility for tasks such as beamforming and single- and multi-user \ac{MIMO}.

\end{abstract}



\section{Introduction}

\IEEEPARstart{N}{ext}-generation wireless systems are expected to provide enhanced performance in terms of capacity, reduced latency, and new functionalities such as integrated sensing and communication \cite{DanAmiShiAlo:20,Pre:J24}. This trend is driving the investigation of fundamental limits using physically consistent models, new technologies, and novel design paradigms to approach them \cite{San:19,BjoEldLarLozPoo:23,YouCaiLiuDiRDumYen:24}.

In this context, the recently introduced \emph{holographic communications} paradigm is envisioned as a holistic way to manipulate, with unprecedented flexibility, the \acf{EM} field generated or sensed by an antenna \cite{DarDec:J21}. It involves designing innovative solutions capable of approaching the fundamental limits imposed by the wireless channel through the massive deployment of \acp{RIS} \cite{DiRDanTre:22}, \ac{ELAA} \cite{WanZhaDuShaAiNiyDeb:23}, and \acp{LIS}, also known as holographic \ac{MIMO} \cite{Dar:J20,DAmSan:23,GonGavJiHuaAleWeiZhaDebPooYue:24}.

In particular, holographic \ac{MIMO} surfaces are envisioned as an efficient implementation of large antenna systems, advancing beyond massive \ac{MIMO} and \acp{ELAA}  using simplified and reconfigurable hardware that is reduced in size, cost and power consumption, and facilitate signal processing in the analog domain \cite{GonGavJiHuaAleWeiZhaDebPooYue:24}.
In holographic \ac{MIMO}, the density of the antenna elements is increased, ideally approaching a continuous distribution, which allows for highly flexible manipulation of \ac{EM} waves. 
Additionally, exploiting higher frequency bands, such as millimeter wave and THz, opens up the potential to utilize the extra \ac{DoF} provided by the channel in the radiating near-field propagation region, even under \ac{LOS} conditions  \cite{PhaIvrGraCreTanNos:18, DecDar:J21,LiuXuWanMuHan:23, ZhaShlGuiDarImaEld:J22, ZhaShlGuiDarEld:J23, TorDecDar:J23}.  
Closely spaced antenna elements (below $\lambda/2$, with $\lambda$ being the wavelength) do not significantly increase the exploitable \ac{DoF} of the channel  \cite{PooTse:15,DAmSan:23}. Nevertheless, when the elements spacing is made very small, strong coupling occurs, and the gain of a linear array shifts from $N$, as in conventional arrays, to $N^2$, being $N$ the number of antenna elements, resulting in the so-called ``superdirectivity" \cite{IvrNos:14}. 
 While achieving very narrow beams using size-limited antennas is highly desirable in many contexts, the superdirectivity effect is only obtained in the end-fire direction of the array and can lead to extremely high driving currents corresponding to high $Q$-factor values (and hence a lower bandwidth) \cite{HanYinMar:22}.

The utilization of high-frequency bands in the millimeter wave and THz ranges, coupled with the integration of antennas with a very large number of elements, is pushing current technology towards insurmountable barriers in hardware complexity and power consumption. This issue, also known as \emph{digital bottleneck}, poses serious challenges for the sustainability of future wireless networks. 
In \ac{MIMO}-based systems, hybrid digital-analog solutions have been extensively explored to partially alleviate these issues by reducing the number of \acf{RF} chains and the digital processing burden, albeit at the expense of flexibility \cite{AlkElALeuHea:14}. 

A promising approach toward sustainability is to delegate part of the signal processing directly to the \ac{EM} level, known as \ac{ESIT} \cite{ZhuWanDaiDebPoo:22,DiRMig:24,BjoChaHeaMarMezSanRusCasJunDem:24}.
 This can be achieved by designing reconfigurable \ac{EM} environments \cite{Dar:J24} using novel \ac{EM} metamaterials devices to perform basic processing functions (e.g., spatial first derivative) \cite{Sil:14}, \acp{RIS}, or the recently introduced \acp{SIM} \cite{AnXuNgAleHuaYueHan:23}.   
However, \acp{SIM} do not fully exploit the possibilities offered by \ac{EM} phenomena in antenna structures for processing and radiation, thereby inheriting some limitations of standard antenna arrays as will be detailed later. 

To overcome the aforementioned issues, we put forth the idea of a \acf{DSA} where  \ac{EM} processing and radiation are performed jointly ``over the air" by using a few active antenna elements surrounded by a cloud of interacting programmable scatterers. 
As will be demonstrated, \acp{DSA} offer several advantages in terms of flexibility and size reduction while minimizing the number of \ac{RF} chains required.  

\subsection{Related State of the Art}

The idea of introducing passive scattering elements close to an antenna to exploit mutual coupling in the reactive near-field region and shape its radiation properties dates back to the early days of wireless transmissions. A pioneering work is that by H. Yagi published in 1928 that introduced the classic Yagi-Uda array antenna. This antenna consists of a set of linear parallel dipoles with a spacing of about $0.2-0.3 \lambda$, where the first dipole is active (i.e., fed with the signal) and the others act as passive scatterers \cite{Poz:97}.
Over the subsequent decades, this idea has been extensively developed in various directions within the antenna theory community.

In more recent times, the introduction of new technologies and materials has enabled the development of new antenna structures designed to meet the demands of higher performance, flexibility, frequency, and lower cost, as required by new-generation wireless systems.   
Among these advancements, metasurface antennas have received particular attention. They are composed of subwavelength elements printed on a grounded dielectric slab. These antennas exploit the interaction between a cylindrical surface wave, excited by a monopole, and an anisotropic impedance boundary condition realized with passive printed elements to produce an almost arbitrary aperture field \cite{FaeMinGonCamMarDelMac:19}. 
The dynamic counterpart to static metasurface antennas is the \ac{DMA}, a type of traveling wave antenna  with reconfigurable subwavelength apertures. Since only one \ac{RF} chain is needed for each row of the planar array, \acp{DMA} offer a compelling solution, balancing flexibility with a reduction in \ac{RF} chains  \cite{ShlAleImaYonSmi:21,ZhaShlGuiDarEld:J23}. 
Static and dynamic metasurface  antennas represent a promising avenue for the research aimed at reducing power consumption in wireless systems, thanks to their ability to shape signals at the \ac{EM} level. 
Other recent technologies include fluid and reconfigurable antennas  \cite{WonNewHaoTonCha:23,PriHarBlaKieKusFriPraMorSmi:04,RodCetJof:14}.
However, they still lack the necessary flexibility to completely replace their digital counterparts, especially for complex linear processing operations.

 An important step toward holographic \ac{MIMO} is the introduction of \acp{SIM} or stacked \ac{RIS} \cite{AnXuNgAleHuaYueHan:23,HasAnDiRDebYue:24,AnDiRDebPooYue:23,AnYueDiRDebPooHan:23}.
 The concept of \ac{SIM} originates from recent advances in optics and microwave circuits designed to perform machine learning tasks using layered \ac{EM} surfaces to mimic a deep neural network for image processing \cite{LiuMaLuo:22,ZeQiaXinWeiTie:24}. 
Similarly, a \ac{SIM} consists of a closed vacuum container with several stacked metasurface layers. The first layer is an active planar antenna array fed by  a set of \ac{RF} chains. The \ac{EM} wave generated by the array travels through subsequent layers composed of numerous reconfigurable meta-atoms. The last layer is in charge of radiating the \ac{EM} field outward.
By properly configuring the transmission properties of each meta-atom in the \ac{SIM}, the system can manipulate the traveling \ac{EM} wave to produce a customized waveform shape. Examples of optimization techniques for \acp{SIM} to achieve various \ac{EM} wave processing functionalities can be found in \cite{AnYueDiRDebPooHan:23,HasAnDiRDebYue:24,AnXuNgAleHuaYueHan:23,AnDiRDebPooYue:23}. Specifically, \cite{AnXuNgAleHuaYueHan:23} investigates a holographic  \ac{MIMO} link where both the transmitter and the receiver use a \ac{SIM}. The authors show that with a 7-layer \ac{SIM} with $\lambda/2$ meta-atom spacing a good fit with the ideal \ac{SVD} precoding and combining tasks is obtained. In \cite{HasAnDiRDebYue:24} and \cite{AnDiRDebPooYue:23}, optimization schemes to realize custom radiation patterns and multi-user downlink beamforming are proposed, whereas the authors in \cite{AnYueDiRDebPooHan:23} demonstrate the  capability to perform a 2D-Fourier  transform in the wave domain.   
 Although an energy consumption model has not yet been studied, it is expected that \acp{SIM} will be more energy efficient compared to conventional digital transceivers.   

The study of \ac{SIM} is still in its infancy, and the main technological and modeling challenges have yet to be fully identified. For instance, the distance between the layers must be several wavelengths to validate the currently adopted cascade model. Power losses and signal distortion caused by the multiple layers and the bounding box require further investigation. A key limitation of \ac{SIM} is that only the final layer radiates, constraining the \ac{EM} wave characteristics to those achievable with a classical planar surface or array, while the role of the hidden layers is to perform an \ac{EM} transformation.

\subsection{Our Contribution}

In this paper, we propose a 3D \ac{EM} antenna structure, termed \ac{DSA}, that offers significant flexibility by providing enhanced processing capabilities directly at the \ac{EM} level. Specifically, a \ac{DSA} consists of a limited number of active antenna elements, each associated with an \ac{RF} chain, surrounded by a cloud of reconfigurable passive scatterers that interact with each other in the reactive near field.

We first characterize the input-output characteristic of the \ac{DSA} as a function of its configuration parameters and provide a fully analytical modeling using Hertzian dipoles.
Subsequently, we propose a strategy to optimize these parameters according to the desired processing functionality and apply this strategy to three different use cases of interest: beamforming, multi-user \ac{MISO}, and \ac{MIMO}. 
Numerical results for each use case demonstrate the great flexibility of the \ac{DSA} in performing various signal processing tasks. For instance, we will show that the tight coupling of \ac{DSA} elements allows for the realization of superdirectivity behavior independently of the beam's direction, in contrast to standard arrays where superdirectivity can only be achieved in the end-fire direction  \cite{IvrNos:10,HanYinMar:22}.    

Compared to \acp{SIM}, in a \ac{DSA}, all the elements contribute jointly to the processing and radiation processes, thereby enhancing flexibility with more compact structures.
Indeed, in a \ac{SIM}, only the elements of the last layer radiate, and thus, the maximum achievable gain is proportional to their number, similar to a conventional antenna array. Conversely, in the numerical results, we demonstrate that a significantly higher gain can be achieved with a \ac{DSA} compared to a standard array.   
Furthermore, \acp{SIM} assume only forward propagation from one layer to the next, with no coupling between the elements of each layer. While this simplifies modeling and optimization significantly, it presents critical challenges in decoupling implementation and imposes size constraints. For example, in \cite{AnXuNgAleHuaYueHan:23} and \cite{HasAnDiRDebYue:24}, a separation of $5\lambda$ between each layer is considered, limiting the possibility of obtaining compact structures. In contrast, our numerical results demonstrate that with a \ac{DSA}, better performance can be achieved with a spacing between elements of $\lambda/4$, thanks to the joint processing and radiation involving all the elements composing the \ac{DSA}.  

The results obtained in this paper demonstrate the potential of \ac{DSA} to realize wave-based computation ``over the air" by leveraging  the coupling between active and scattering elements, thus minimizing the number of \ac{RF} chains and baseband processing in the digital domain. This is achieved through a semi-passive structure with minimal energy consumption and nearly zero latency, as the processing occurs at the speed of light.

\begin{figure*}[!t]
\centering 
\centering\includegraphics[width=2\columnwidth]{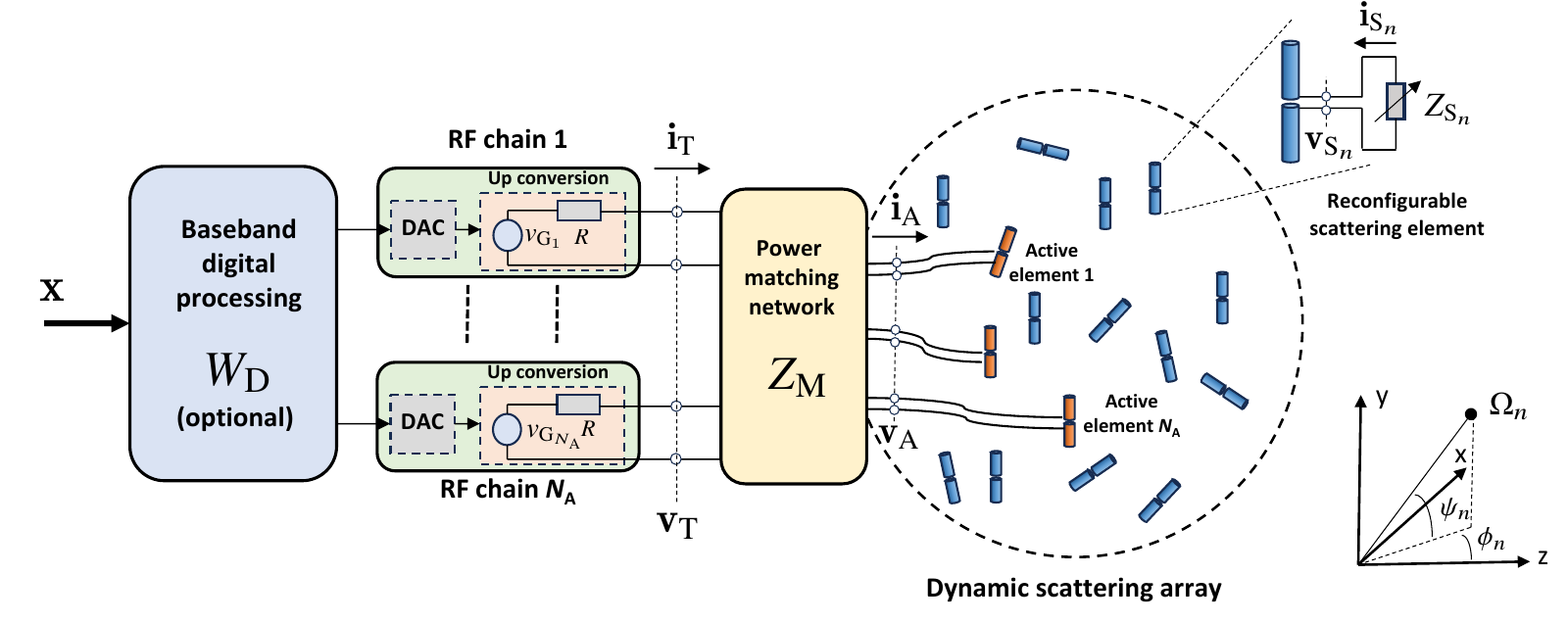}
\caption{Principle scheme of a dynamic scattering array.} 
\label{Fig:System}
\end{figure*}

\subsection{Notation and Definitions}
Lowercase bold variables denote vectors in the 3D space, i.e.,  $\boldp=\versorx \, p_x + \versory \, p_y + \versorz \, p_z$
 is a vector with cartesian coordinates $(p_x, p_y, p_z)$, 
 $\versorp$ is a unit vector denoting its direction, and  $p=|\boldp|$ denotes its magnitude, where $\versorx$, $\versory$, and $\versorz$ represent the unit vectors in the $x$, $y$ and $z$ directions, respectively.  
 The scalar product between vectors $\boldp$ and 
 $\boldr$ is indicated with ${\boldp} \scalprod {\boldr}$. 
 $\delta(x)$ represents  the Dirac delta distribution and $\delta(\boldp)=\delta(p_x)\, \delta(p_y)\, \delta(p_z)$ its 3D version.  
Boldface capital letters are matrices (e.g., $\boldA$), where $\boldI_N$ is the identity matrix of size $N$, $\boldzero_N$ is the zero matrix of size $N$. With reference to a generic matrix $\boldA$,  $a_{n,m}=[\boldA]_{n,m}$ represents its $(n,m)$th element, whereas  $\boldA\transpose$ and $\boldA\ctranspose$ indicate the transpose and the conjugate transpose of $\boldA$. We denote with $\boldA^{\dag}=(\boldA\ctranspose \, \boldA)^{-1} \boldA\ctranspose$ the Moore-Penrose inverse (pseudo-inverse) and with $\left \|  \boldA  \right \|_{\text{F}} $ the Frobenius norm of matrix $\boldA$. We indicate with $\Real{z}$, $\Imag{z}$, and $z^*$, respectively,  the real part,  imaginary part, and complex conjugate of the complex number $z$, and $\jmath$ the imaginary unit.
Denote with $\eta=377\,$Ohm the free-space impedance and $c$  the speed of light.

\subsection{Paper Organization}
The rest of the paper is organized as follows: 
Sec.~\ref{Sec:Model} introduces and models the proposed \ac{DSA}.
In Sec.~\ref{Sec:Dipoles}, an analytical characterization of the \ac{DSA} is developed using Hertzian dipoles. The optimization of the parameters characterizing the \ac{DSA} is addressed in general in Sec.~\ref{Sec:Optimization}, and further declined in 3 use cases. Numerical results associated with the 3 use cases are provided in Sec.~\ref{Sec:NumericalResults}. Finally, the conclusions are drawn in Sec.~\ref{Sec:Conclusion}.

\section{\acf{DSA} Modeling}
\label{Sec:Model}

We consider the transmitting system illustrated in Fig.~\ref{Fig:System}, where the \ac{DSA} is composed of $\Na$ active antenna elements surrounded by a cloud of $\Ns$ reconfigurable passive scatterers. The $n$th element (active or passive) of the \ac{DSA}, with $n=1,2, \ldots N$, $N=\Na+\Ns$, is located in position $\boldp_n$. The $n$th scatterer, with $n=1,2, \ldots \Ns$, is itself an antenna element terminated by a reconfigurable impedance load $\Zs{n}$. To avoid power dispersion, we impose that  the load is purely reactive, i.e., $\Zs{n}=\jmath \theta_n$, where $\theta_n \in \mathbb{R}$ is the reconfigurable reactance.\footnote{From the technological point of view, the reconfigurable impedance load can be implemented using solutions similar to those applied for \acp{RIS} such as PIN diodes and varactors \cite{AlePha:23}.} 
Denote with  $\btheta=\left [\theta_1, \theta_2, \ldots, \theta_{\Ns} \right ]\transpose$ the set of reconfigurable reactances and with $\bZs=\diag{\Zs{1}, \Zs{2}, \ldots , \Zs{\Ns}} $ the corresponding set of loads.  
 
The \ac{DSA} can be modeled as a linear $N$-port network. 
The real-value bandpass voltage $v^{\text{(RF)}}(t)$ present at the generic port of the network is a modulated \ac{RF} signal with carrier frequency $f_0$ and it can be represented as 
\begin{equation} \label{eq:vrf}
	v^{\text{(RF)}}(t)=\sqrt{2} \Real{v(t) \, e^{\jmath 2\pi f_0\,  t} }
\end{equation}
where $v(t)$ is the associated complex voltage envelope. 
Similarly, we define the complex envelope $i(t)$ of the real-value bandpass current $i^{\text{(RF)}}(t)$ flowing into the port.  Since the modulated signal carries information, then they have to be treated as random processes. The (average) active power delivered to the port is $P=\EX{\Real{v^{\text{(RF)}}(t) \, i^{\text{(RF)}}(t)}}=\EX{\Real{v^*(t) \, i(t)}}$, where $\EX{\cdot}$ is the statistical expectation and the last equality holds thanks to the arbitrary constant $\sqrt{2}$ in \eqref{eq:vrf}.
We assume that the bandwidth $B$ of $v^{\text{(RF)}}(t)$ is sufficiently small such that the network properties are constant within $B$ and can be evaluated at the carrier frequency $f_0$.

In a compact description, we define  $\bva \in \mathbb{C}^{\Na \times 1}$ and $\bia \in \mathbb{C}^{\Na \times 1}$ the complex voltage and current envelopes (in the following denoted simply as voltages and currents) at the ports of the $\Na$ active antennas (i.e., the input ports of the \ac{DSA}), and $\bvs \in \mathbb{C}^{\Ns \times 1}$ and $\bis \in \mathbb{C}^{\Ns \times 1}$ the voltages and currents at the ports of the $\Ns$ scatterers.\footnote{For notation convenience, in the following we omit the time dependence. }
We collect all the currents and voltages at the $N$ ports of the \ac{DSA} in the vectors 
  \begin{align} 
&\bi=\left [
\begin{array}{c}
 \bia      \\
 \bis          
\end{array}
   \right ]   
 &\bv=\left [
\begin{array}{c}
 \bva      \\
 \bvs          
\end{array}
   \right ] \, .
\end{align}

All the interactions between the elements of the \ac{DSA} are captured by the impedance matrix $\bZ  \in \mathbb{C}^{2N \times 2N}$, which does not depend on the reconfigurable loads, and relates the voltages and currents of the $N$ ports as $\bv=\bZ \, \bi$.
In particular, the $(n,m)$th element of $\bZ$ represents the mutual coupling coefficient between the $n$th and $m$th elements  obtained as the ratio between the open-circuit voltage observed at the $m$th port and the excitation current applied to the $n$th port supposing that the remaining ports are kept open (no current flow) \cite{BalB:16}.   It is convenient to divide the impedance matrix  $\bZ$ into the sub-matrices $\bZaa$, $\bZas$, $\bZsa$, and $\bZss$ so that we can write
\begin{align} \label{eq:vavs}
\left [
\begin{array}{c}
 \bva      \\
 \bvs          
\end{array}
   \right ]=\left [
\begin{array}{cc}
 \bZaa & \bZas     \\
 \bZsa & \bZss          
\end{array}
   \right ] \cdot \left [
\begin{array}{c}
 \bia      \\
 \bis          
\end{array}
   \right ] \, .
\end{align}

The voltages $\bva$ and currents $\bia$ at the input port of the \ac{DSA} are related by the input impedance of the \ac{DSA} $\bZa \in \mathbb{C}^{\Na \times \Na}$ as $\bva=\bZa \, \bia$. 
By elaborating \eqref{eq:vavs} and considering that at the scatterers' ports it is $\bvs=-\bZs \, \bis$, we obtain
\begin{align} \label{eq:bZa}
\bZa=\bZaa-\bZas \, \left (\bZss+\bZs  \right )^{-1} \, \bZsa 
\end{align}
which depends on parameters $\btheta$ through the scatterers' loads $\bZs$.

The $\Na$ active elements are connected to $\Na$ \ac{RF} chains, including \ac{DAC} and up-conversion stages, through a $2\Na$-port power matching network whose purpose is to maximize the power transfer between the \ac{RF} chains and the \ac{DSA}. 
The relationship between the voltages and the currents present at the ports of the matching network is given by  
\begin{align} \label{eq:vtva}
\left [
\begin{array}{c}
 \bvt      \\
 \bva          
\end{array}
   \right ]=\bZm \, \left [
\begin{array}{c}
 \bit      \\
 -\bia          
\end{array}
   \right ]
\end{align}
where, assuming no losses in the power matching network, $\bZm$ is equal to  \cite{IvrNos:10}
\begin{align} \label{eq:bZm}
\bZm=\left [
\begin{array}{cc}
 \boldzero_{\Na} & -\jmath \sqrt{R} \, \Real{\bZa}^{\frac{1}{2}}      \\
 -\jmath \sqrt{R} \, \Real{\bZa}^{\frac{1}{2}} & -\jmath \Imag{\bZa}           
\end{array}
   \right ] 
\end{align}
where $R$ is the output resistance of each \ac{RF} chain.
As it can be noticed from \eqref{eq:bZm}, the matching network depends on the \ac{DSA} input impedance $\bZa$ and hence on the parameters $\btheta$. Therefore, for a given \ac{DSA} configuration, the matching network should be changed accordingly. This is possible using, for instance, adaptive matching networks \cite{AliShu:20}. In case a fixed network is employed instead, then not all the available power would be transferred to the \ac{DSA}. 
With the power matching network, it turns out that $\bvt=\bZt \, \bit$, with $\bZt=R\, \boldI_{\Na}$ and, assuming lossless antenna elements, the transmitted power corresponds to the radiated power $\Prad$ given by
\begin{align}
	\Ptx=\Prad=\EX{\bia\ctranspose \, \Real{\bZa} \, \bia}=\frac{\EX{\bvt\ctranspose \, \bvt}}{R} \, .
\end{align}
The reactive power can be evaluated as well as 
\begin{align}
	\Preact=\EX{\bia\ctranspose \, \Imag{\bZa} \, \bia} 
\end{align}
from which the $Q$-factor of the system can be obtained as $Q=\Preact/\Prad$. The inverse of the $Q$-factor can be used as a rough estimate of the bandwidth of the \ac{DSA} when $Q\gg 1$\cite{BalB:16}. 

We also foresee an optional baseband linear processing block implemented in the digital domain (digital precoder), described by the matrix $\bWd \in \mathbb{C}^{\Na \times \Na}$, whose input is represented by the information vector $\boldx \in \mathcal{C}^{\Na \times 1}$ to be transmitted. The matrix $\bWd$ 
relates the transmitted information vector $\boldx$ and the voltages $\bvt$ at the input of the matching network, i.e., $\bvt=\bWd\, \boldx$.  In the following, we impose the arbitrarily normalization $\| \bWd \|_{\text{F}}=\sqrt{R\, \Na}$. Such a normalization ensures that $\EX{\boldx\ctranspose \, \boldx}=\Ptx$, as typically considered in the signal processing community.
The reconfigurable reactances $\btheta$ along with the elements of matrix $\bWd$ represent the set of parameters to be optimized to obtain the desired processing functionality, as will be described later.

From \eqref{eq:vtva} and \eqref{eq:bZm} it is
\begin{align} \label{eq:bia}
\bia=\frac{1}{\jmath \sqrt{R}} \, \Real{\bZa}^{-\frac{1}{2}}   \bvt =\boldM(\btheta) \, \bvt \, .
\end{align}
Moreover, we have
\begin{align} \label{eq:bi}
\bi
=\left [
\begin{array}{c}
 \boldI_{\Na}    \\
 -\left ( \bZss + \bZs \right )^{-1} \bZsa        
\end{array}
   \right ] \cdot 
  \bia =\boldC(\btheta)\, \bia \, .
\end{align}

By combining \eqref{eq:bia} and \eqref{eq:bi}, we obtain a compact relationship between the information vector $\boldx$ and the total current $\bi$ flowing in the \ac{DSA}
\begin{equation}
\bi=\boldC(\btheta) \, \boldM(\btheta) \, \bWd\, \boldx= \bWem \, \bWd \, \boldx 
\end{equation}
where $\bWem$ 
\begin{align} \label{eq:Wa}
\bWem
=\frac{1}{\jmath \sqrt{R}} \, \left [
\begin{array}{c}
  \Real{\bZa}^{-\frac{1}{2}}    \\
 - \left ( \bZss + \bZs \right )^{-1} \bZsa   \, \Real{\bZa}^{-\frac{1}{2}}      
\end{array}
   \right ] 
\end{align}
accounts for the \ac{EM}-level signal processing operated by the reconfigurable \ac{DSA} as a function of the parameters $\btheta$.

Consider now $K$ test points located in positions $\boldt_k$, with $k=1,2, \ldots , K$, and  that in each test point a conventional receiving antenna is used to receive the signal with gain $\Gr$. 
As commonly done in the literature, we assume that  $|\boldt_k-\boldp_n|\gg \lambda$, $\forall k,n$, that is, the test points are located in the radiative region of the antenna structure and the receiving antennas do not affect the transmitting \ac{DSA}. 
As an example, the set of $K$ antennas could represent a conventional receiving antenna array of a \ac{MIMO} system.  
The useful component (i.e., without noise) of the received signal at the test positions is 
\begin{align} \label{eq:by}
	\boldy=\bHc \, \bi=\bHc\, \bWem \, \bWd \, \boldx 
	 =\boldH(\btheta, \bWd) \, \boldx
\end{align}
where $\bHc$ is the transimpedance matrix of the radio channel accounting for the propagation effects and $\boldH(\btheta, \bWd)$ is the end-to-end baseband equivalent channel matrix as commonly defined in signal processing.

\section{DSA with Hertzian Dipoles}
\label{Sec:Dipoles}

To further develop our framework and derive some numerical examples from which obtaining important insights about the potential of \acp{DSA}, we suppose that all the antenna elements in the \ac{DSA}, both active and scatterers, are modeled as Hertzian dipoles.
In addition to making the derivation analytically tractable, the consideration of Hertzian dipoles can also be justified by the fact that the discrete infinitesimal dipole approximation is a flexible technique in computational electromagnetics for modeling and solving more complex antenna/scatterers structures \cite{MikKis:07}. 

Accordingly, we model the $n$th dipole as a very short wire of length $l_n$ and radius $r_n$, with $l_n, r_n \ll \lambda$ and generic orientation $\bOmega_n=\left [ \sin (\phi_n) \, \cos (\psi_n) \, , \sin (\phi_n) \, \sin (\psi_n) \, , \cos (\phi_n) \right ]\transpose$, where $\phi_n$ and $\psi_n$ represent, respectively, the elevation angle and the angle with respect to axis $z$, according to the reference coordinates system illustrated in Fig. \ref{Fig:System}. 
The $n$th element $i_n$ of vector $\bi$ is  the current present at the port of the $n$th dipole. 

The total current density distribution in the \ac{DSA} is given by
\begin{align} \label{eq:jm}
  \Jm(\boldp) =\sum_{n=1}^N  \bOmega_n \, l_n\, i_n \, \delta(\boldp-\boldp_n) \, .
\end{align} 
%
Therefore, the \ac{EM} field at the generic location $\boldp$ excited by the current density  $\Jm(\boldp)$ in free space can be evaluated as \cite{Dar:J24} 
\begin{align} \label{eq:Em}
 \Em(\boldp)=  \int_{\mathcal{V}} \Gej{\boldp-\bolds} \,  \Jm(\bolds) \, d \bolds 
\end{align}
where  $\mathcal{V}$ is the volume in which the current flows and $\Gej{\boldr}$ is the Green's dyadic given by \cite{BalB:16}
%
%
\begin{align} \label{eq:Gej}
& \Gej{\boldr} = -\jmath \frac{ \eta \, e^{-\jmath \kappazero \, |\boldr|}}{2 \lambda |\boldr|}  \nonumber \\
& \quad \quad \cdot \left [\left ( \boldI - \hat{\boldr} \, \hat{\boldr}\transpose \right ) + \frac{\jmath \lambda }{2 \pi |\boldr|} \left  ( \boldI - 3 \hat{\boldr} \, \hat{\boldr}\transpose \right ) - \frac{ \lambda^2 }{(2 \pi |\boldr|)^2}  \left ( \boldI - 3 \hat{\boldr} \, \hat{\boldr}\transpose \right )\right ] 
\end{align}
with $\kappazero=2 \pi /\lambda$ being the wave number and where we grouped the terms multiplying the factors $1/|\boldr|$, $1/|\boldr|^2$, and $1/|\boldr|^3$. 
By substituting \eqref{eq:jm} in \eqref{eq:Em}, we obtain
\begin{align} \label{eq:Em1}
 \Em(\boldp)=  \sum_{n=1}^N   \, l_n\, i_n \,  \Gej{\boldp-\boldp_n} \, \bOmega_n \, .
\end{align}

The open-circuit voltage at the $m$th dipole caused by the current $i_n$ of the $n$th dipole by assuming all the other currents are equal to zero is
\begin{align} \label{eq:V0m}
 V_{m}^{(\text{oc})}=    l_m \, l_n\, i_n \,  \bOmega_m\transpose\,  \Gej{\boldp_m-\boldp_n} \, \bOmega_n \, .
\end{align}
From \eqref{eq:V0m}, the generic element $Z_{m,n}=[\bZ]_{m,n}$ of the impedance matrix $\bZ$  associated with the \ac{DSA}  can be easily obtained as 
\begin{align}
Z_{m,n}=\frac{V_{m}^{(\text{oc})}}{i_n}  =l_n \, l_m \, \bOmega_n\transpose \, \Gej{\boldp_n-\boldp_m} \, \bOmega_m 
\end{align}
for $n \neq m$. For $n=m$,  $Z_{n,n}$ corresponds to the self-impedance of the Hertzian dipole which is given by  \cite{BalB:16}
\begin{align}
	Z_{n,n}=\frac{2}{3} \pi \eta \left (\frac{l_n}{\lambda} \right )^2 + \frac{ \ln \left (  \frac{l_n}{r_n}   \right )}{\jmath f_0 \pi^2 \epsilon l_n}  \, .
\end{align}

In the free-space propagation condition, the channel transimpedance matrix $\bHc$ in \eqref{eq:by} can be easily derived. 
To elaborate, the electrical field intensity at the $k$ receiving test antenna located at $\boldt_k$ can be computed once the currents $\bi$, flowing into the dipoles (active and scatterers), are known 
\begin{align}
e_k=\bOmegak\transpose \, \Em(\boldt_k)=\sum_{n_1}^N  l_n \, \bOmegak\transpose \, \Gej{\boldt_k-\boldp_n} \, \bOmega_n \, i_n
\end{align}
 where  $\bOmegak$ is the polarization direction of the $k$th receiving antenna. As a consequence, the received signal (assuming matched load and normalized to an impedance $R_{\text{L}}=1\,$Ohm) is given by $y_k= e_k \sqrt{\frac{A_{\text{eff}} R_{\text{L}}}{\eta}}= e_k \sqrt{ \frac{\lambda^2 \, \Gr }{4 \pi \eta}}$. 
 Finally, the $(k,n)$th element of the transimpedance matrix $\bHc$ is given by
 \begin{equation} \label{eq:Rc1}
 [\Hc]_{k,n}=\sqrt{ \frac{\lambda^2 \, \Gr }{4 \pi \eta}} l_n \, \bOmegak\transpose \, \Gej{\boldt_k-\boldp_n} \, \bOmega_n \, .
 \end{equation}
Since it is $|\boldr_{k,n}|=|\boldt_k-\boldp_n| \gg \lambda$, \eqref{eq:Rc1} can be well approximated by neglecting the $1/|\boldr|^2$ and $1/|\boldr|^3$ components (reactive field) in \eqref{eq:Gej} thus leading to 
\begin{align}
\Gej{\boldr} \simeq -\jmath \frac{\eta \kappazero}{4 \pi  |\boldr|}     \, e^{-\jmath \kappazero \, |\boldr|} \, \left ( \boldI_3 - \hat{\boldr}_{k,n} \, \hat{\boldr}\transpose_{k,n} \right )
\end{align}
and
\begin{align}
 [\Hc]_{k,n}  
 \simeq -\jmath \frac{\eta}{2  \lambda |\boldr|}      \sqrt{ \frac{ \Gr }{4 \pi \eta}} l_n  \, e^{-\jmath \kappazero \, |\boldr|}  \bOmegak\transpose  \left ( \boldI_3 - \hat{\boldr}_{k,n} \, \hat{\boldr}\transpose_{k,n} \right ) \bOmega_n \, .
 \end{align}


%
%

\section{DSA Optimization}
\label{Sec:Optimization}

In this section, we illustrate a possible optimization strategy of the \ac{DSA} along with its application to 3 different use cases. Numerical results associated with each use case will be reported in Sec.~\ref{Sec:NumericalResults}.

\subsection{Joint Optimization of the DSA and the Digital Precoder}

For a given propagation scenario characterized by the transimpedance matrix $\bHc$ and $K$ test points, suppose we fix a specification $\boldHopt$ on the desired end-to-end channel matrix. The reconfigurable  parameters of the \ac{DSA} and the digital precoder can be obtained by solving the following constrained optimization problem:
\begin{align}
\label{eq:opt}
& 
\minimize{\btheta, \, \bWd, \,  \alpha}  \left \|  \alpha \, \bHc \, \bWem \, \bWd - \boldHopt  \right \|_{\text{F}} \\
& s.t. \quad \| \bWd \|_{\text{F}}^2=R\, \Na  	\nonumber
\end{align}
where the scalar $\alpha \in \mathbb{R}$ accounts for the possible lack in the link budget to achieve $\boldHopt$ that has to be eventually coped with increased transmitted power. 

Unfortunately, in general, the constrained optimization problem in \eqref{eq:opt} is not convex thus making its numerical solution more challenging. A possible approach to translating it into an unconstrained optimization problem is to resort to the following alternate optimization procedure:

\begin{itemize}
\item {STEP 0:}
We set the candidate vector $\hat{\btheta}$  to an initial guess value, for instance randomly chosen or equal to zero.

\item {STEP 1:}

For a fixed $\hat{\btheta}$, it is possible to find in closed form the values of $\bWd$ and $\alpha$ that minimize  the objective function   and satisfy the constraint in \eqref{eq:opt}. In particular, we solve the following equation
\begin{align}
\alpha \, \bHc \, \bWemhat \, \bWd = \boldHopt
\end{align}
of which the minimum norm solution is
\begin{align}
 \alpha \bWd  =  \left (\bHc\,  \bWemhat    \right )^{\dag} \boldHopt \, .
\end{align}

The corresponding values of $\hat{\alpha}$ and $\bWdhat$ are, therefore,
\begin{align} \label{eq:hatalpha}
 \hat{\alpha}   = \frac{1}{\sqrt{R\, \Na} } \left \| \left (\bHc\,  \bWemhat    \right )^{\dag} \boldHopt \right \|_{\text{F}}
\end{align}
and
\begin{align}
  \bWdhat  = \frac{1}{\hat{\alpha} } \left (\bHc\,  \bWemhat    \right )^{\dag} \boldHopt \, .
\end{align}

\item {STEP 2:} 

Starting from the candidates  $\hat{\alpha}$ and $\bWdhat$ from STEP 1, the following unconstrained minimization problem involving only $\btheta$ is solved numerically
\begin{align}
\label{eq:opt1}
& \hat{\btheta} =\argmin{\btheta}  \left \|  \hat{\alpha} \, \bHc \, \bWemhat \, \bWd - \boldHopt  \right \|_{\text{F}}  \, .
\end{align}


\item STEP 3: Repeat STEPS 1 and 2 for a given number $N_{\text{alt}}$ iterations or until no significant variations of the parameters take place.
\end{itemize} 

The inclusion of the digital precoder, while not strictly necessary, may aid in the optimization process of the \ac{DSA} by handling a portion of the processing when $\Na>1$. If omitted, we set $\bWdhat=\bWd=\sqrt{R} \, \boldI$, and only proceed with STEP 2 once.
Of course, alternative optimization problems apart from \eqref{eq:opt} can also be considered, such as those involving the $Q$-factor as an optimization target or constraint.
In many applications (e.g., beam steering, as seen in use case 1 below), the optimization process can be performed offline, computed once to create a dictionary of various \ac{DSA} parameters for different steering directions of interest or, more broadly, different beam forms. In such cases, the complexity of the optimization algorithm is not a concern.
However, in other scenarios, performing optimization offline may not be feasible, necessitating online optimization based, for example, on the \ac{CSI}. In this situation, developing efficient algorithms to minimize \eqref{eq:opt1}, potentially considering the structure of $\bWemhat$, becomes pertinent. This aspect exceeds the scope of this paper and is reserved for future investigations.
In the numerical results, the standard numerical tool based on the quasi-Newton method \cite{FleB:80} with $N_{\text{i}}$ iterations is utilized to minimize \eqref{eq:opt1}.

\begin{figure}[!t]
\centering 
\centering\includegraphics[width=1\columnwidth]{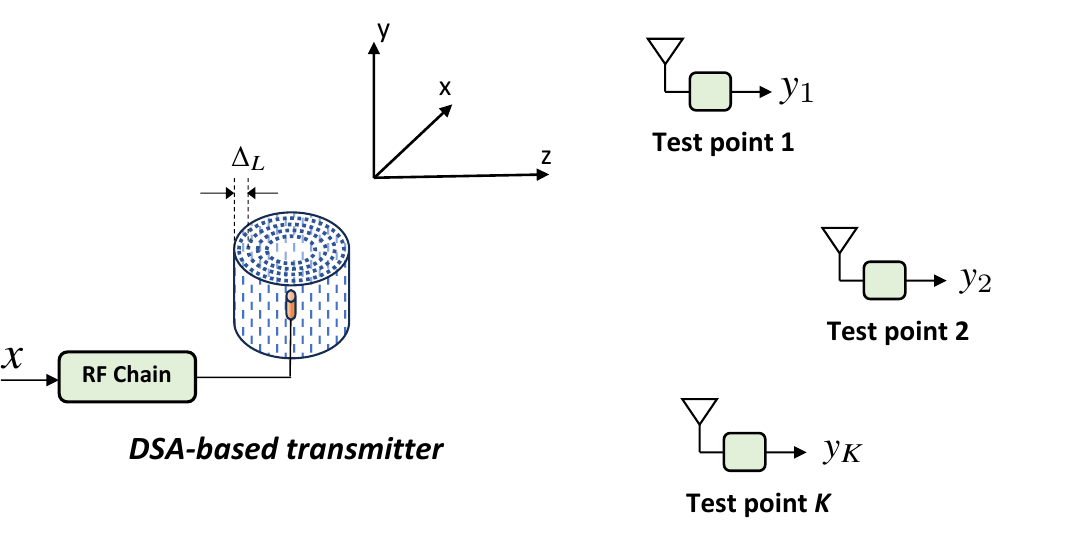}
\caption{Use case 1: Single RF chain DSA for high gain beamforming.} 
\label{Fig:Example1}
\end{figure}

\subsection{Use Case 1: Beam Forming}
\label{Sec:Example1}

Suppose we want to design a single-input \ac{DSA} ($\Na=1$, no digital precoding) realizing a generic radiation diagram, for instance, imposing that at the test locations $\boldt_k$,  the channel matrix $[\boldHopt]_{k,1}$, $k=1,2, \ldots, K$, assumes the desired value (see Fig.~\ref{Fig:Example1}). 
For instance, one might define a uniform set of test points $\boldt_k=[d\sin \phi_k ,0 ,d \cos \phi_k]\transpose$ on the $x-z$ plane a distance $d$ in the far-field region of the \ac{DSA}, with $\phi_k=2 \pi k / K$, and maximize the radiation diagram at the $k=\tilde{k}$th direction $\phi_{\tilde{k}}$ (beam steering). This can be achieved by setting   $[\boldHopt]_{k,1}=\sqrt{\Prx}$, for $k=\tilde{k}$, and zero otherwise, being $\Prx$ the desired received power.  Incidentally, after the optimization process, the resulting parameter $\hat{\alpha}^2$ will represent the required transmitted power $\Ptx$ to satisfy the link budget.  
It is worth noticing that, in principle, only the optimization at the single test point $\boldt_{\tilde{k}}$, i.e., $K=1$, would suffice to obtain the desired beam steering result. However, the addition of many other test points set to zero helps the minimization algorithm to speed up and avoid the convergence to bad local minima.   

\begin{figure}[!t]
\centering 
\centering\includegraphics[width=1\columnwidth]{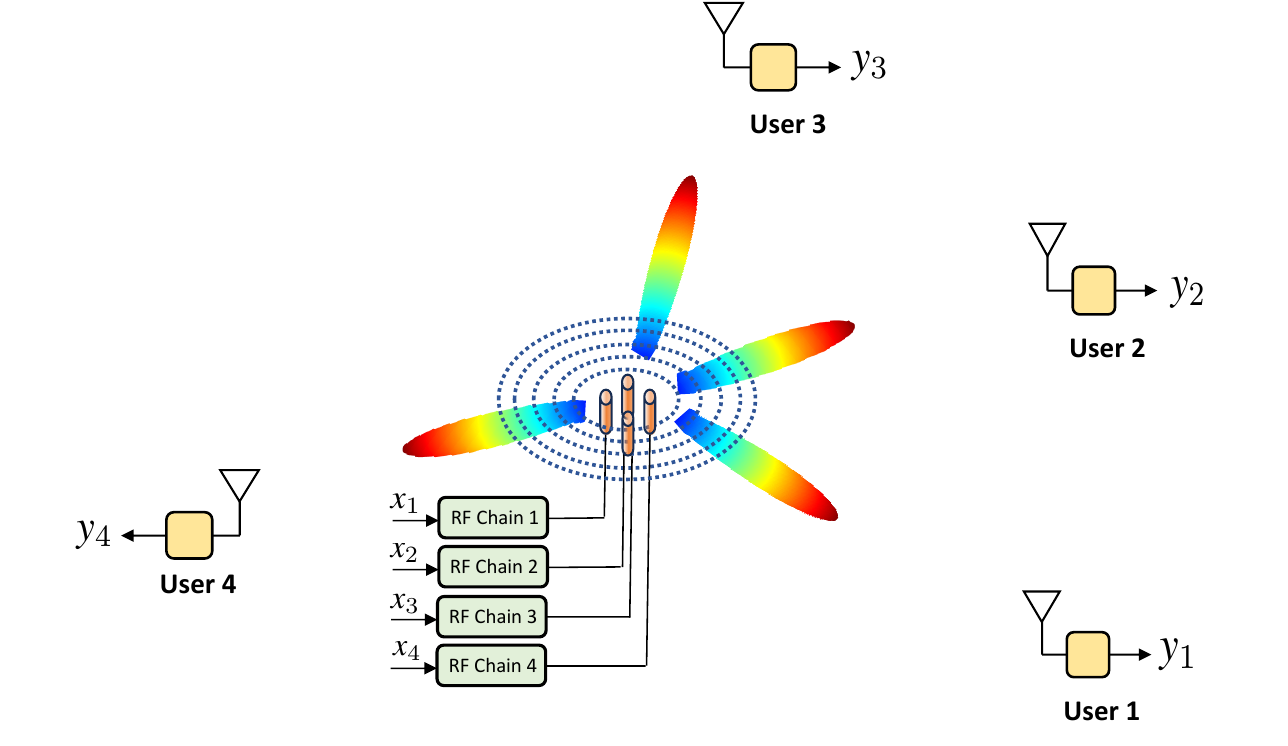}
\caption{Use case 2: Multi-user downlink MISO with a DSA.} 
\label{Fig:Example2}
\end{figure}

\subsection{Use Case 2: Multi-user \ac{MISO}} 
\label{Sec:Example2}

In this second use case, we consider a multi-user \ac{MISO} downlink scenario shown in Fig.~\ref{Fig:Example2},  where an  \ac{DSA} with  $\Na$ \ac{RF} chains (one per user) has to be designed to serve simultaneously $K=\Na$ single-antenna users located in generic positions $\boldt_k$, $k=1,2, \ldots, K$, by maximizing the \ac{SINR} at each user according to the zero-forcing criterium \cite{TseVis:B05}. Users can be both in the radiating near-field and far-field regions of the \ac{DSA}.

Given the channel transimpedance matrix $\bHc=\left [\bhc_1 ; \bhc_2 ; \ldots ; \bhc_{\Na}  \right ]$, where $\bhc_k$ is the $N \times 1$ channel  vector associated with the $k$th user, the precoding matrix is $\boldV=\beta \, \bHc^{\dag}$, where $\beta$ is a normalization factor to make $\boldV$ unitary, i.e., $ \boldV\ctranspose \boldV =\boldI_N$ \cite{TseVis:B05}. Therefore, the \ac{DSA} and the digital precoder must be designed such that
 \begin{equation} \label{eq:ZF}
 	 \bWemhat \, \bWd=\boldV=\beta \, \bHc^{\dag} 
 \end{equation}
 leading to
 \begin{equation}  
\boldy=\bHc \,  \bWemhat \, \bWd \, \boldx=\bHc \, \bHc^{\dag} \, \boldx \, .
\end{equation}

This is equivalent to setting $\boldHopt=\bHc \, \bHc^{\dag}$ in the optimization problem in \eqref{eq:opt}.

\subsection{Use Case 3: Multi-layer \ac{MIMO} Communication} 
\label{Sec:Example3}

\begin{figure}[!t]
\centering 
\centering\includegraphics[width=1\columnwidth]{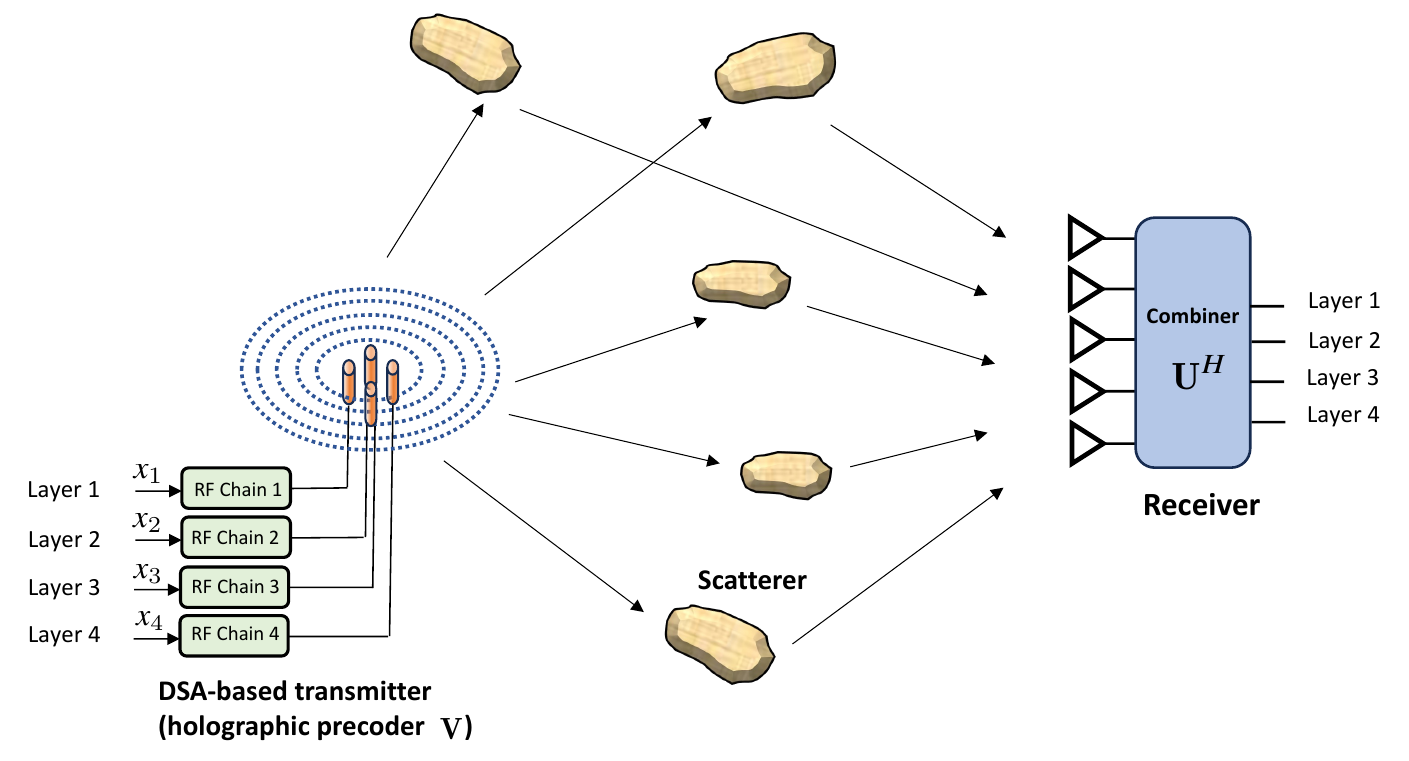}
\caption{Use case 3: Multi-layer MIMO with the DSA acting as an holographic precoder.} 
\label{Fig:Example3}
\end{figure}

In this use case, we consider a \ac{MIMO} communication with a receiving conventional \ac{ULA} with $K$ elements spaced at $\lambda/2$ illustrated in Fig.~\ref{Fig:Example3}. The purpose is to design the \ac{DSA} in such a way an optimal multi-layer \ac{MIMO} link is established between the transmitter and the receiver. It is well known from \ac{MIMO} theory that for a given generic propagation scenario with channel transimpedance matrix $\bHc$ characterized by rank $r$, up to $r$ parallel orthogonal links (layers) can be established between the transmitter and the receiver on which independent data stream can be transmitted \cite{TseVis:B05}. To exploit all of them, it must be $\Na=r$ and the \ac{DSA} must implement a suitable precoding strategy, i.e., act as an \emph{Holographic \ac{EM} precoder}.  
  
Let us introduce the \ac{SVD} of the channel transimpedance
 \begin{equation}
 	\bHc=\boldU\, \bLambda\,  \boldV\ctranspose
 \end{equation}
 where $\boldU$ and $\boldV$ contain the left and right eigenvectors, respectively, and $\bLambda$ is a diagonal matrix gathering the singular values of the channel transimpedance matrix $\bHc$.
 A  typical \ac{MIMO} setup, assuming the \ac{CSI} is available at the transmitter, requires a precoding operation $\boldV$ at the transmitter and a combining operation $\boldU\ctranspose$ at the receiver such that the channel is diagonalized, i.e., the end-to-end channel matrix is proportional to the diagonal matrix $\bLambda$. Assuming the receiver performs the combining operation $\tilde{\boldy}=\boldU\ctranspose\, \boldy$, the \ac{MIMO} channel is diagonalized if we design the \ac{DSA} implementing the precoding $\boldV$ by setting 
  \begin{equation} \label{eq:SVD}
 	 \bWemhat \, \bWd=\boldV
 \end{equation}
 or, equivalently,
 \begin{equation}
 	\boldHopt=\boldU\ctranspose \, \bLambda \, .
 \end{equation}
 In this manner, it results
 \begin{equation}
 	\tilde{\boldy}= \boldU\ctranspose \boldH(\btheta, \bWd) \, \boldx=\boldU\ctranspose \bHc \,  \bWemhat \, \bWd \, \boldx= \bLambda \,  \boldV\ctranspose \,  \boldV \, \boldx=\bLambda \, \boldx
 \end{equation}
and $\Na$ parallel communication layers can be established. The designed \ac{DSA} implements the optimal precoding by using no more than $\Na=r$ RF chains (i.e., the minimum possible value) which is much simpler and less energy consuming than any conventional full digital or hybrid solution.  

In our numerical investigations, we discovered that a more robust optimization is obtained by solving the following equivalent problem instead of \eqref{eq:opt1}
   \begin{align}
\label{eq:opt2}
& \hat{\btheta} =\argmin{\btheta}  \left \|  \hat{\alpha} \, \boldU\ctranspose \bHc \, \bWemhat \, \bWd - \bLambda  \right \|_{\text{F}}  
\end{align}
 that consists in multiplying both terms in \eqref{eq:opt1} by $\boldU\ctranspose$.

\section{Numerical Results}
\label{Sec:NumericalResults}

In this section, we present some numerical results related to the use cases described in  Sec.~\ref{Sec:Optimization} with the purpose of validating and investigating the potential of the proposed \ac{DSA} with respect to classical array structures. 
The following parameters have been considered in the numerical evaluations if not otherwise specified: $f_0=28\,$GHz, $\Gr=0\,$dB, $R=50\,$Ohm, $l_n=\lambda/50$, $r_n=10^{-2} \lambda$, $\forall n$, vertical polarization ($\bOmega_n=[0,1,0]\transpose$, $\forall n$).
  
\begin{figure}[!t]
\centering 
\centering\includegraphics[width=0.9\columnwidth]{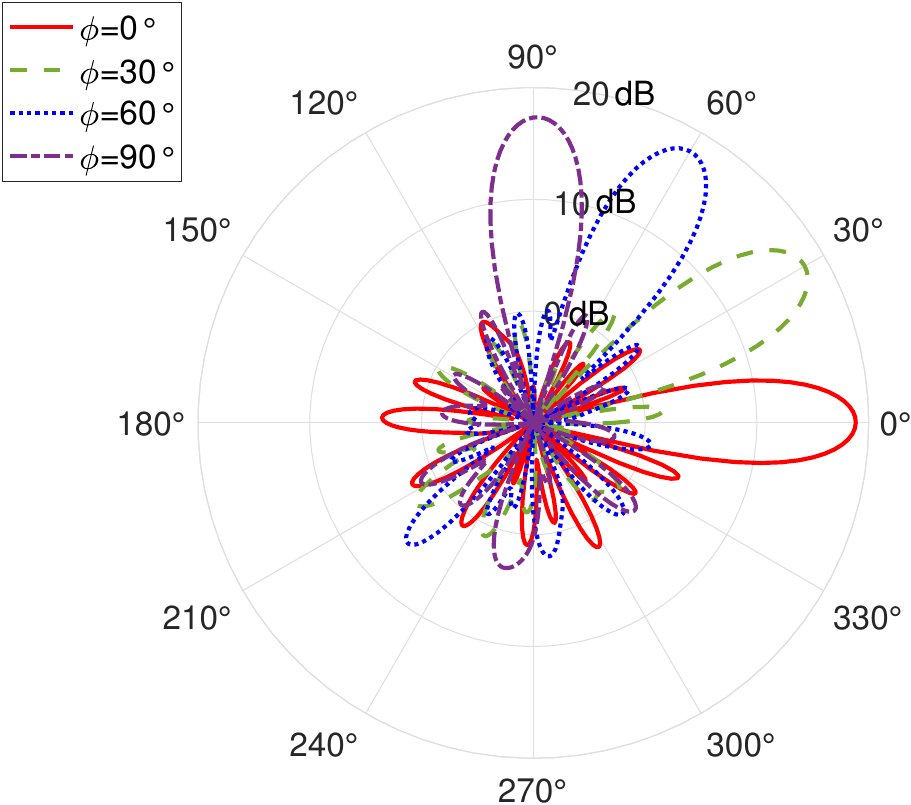}
\caption{Radiation diagram of a disk-shape DSA ($L_{\text{R}}=1$) with $\Delta_L$ and $L=5$.} 
\label{Fig:PatternDSA}
\end{figure}

\begin{figure}[!t]
\centering 
\centering\includegraphics[width=0.9\columnwidth]{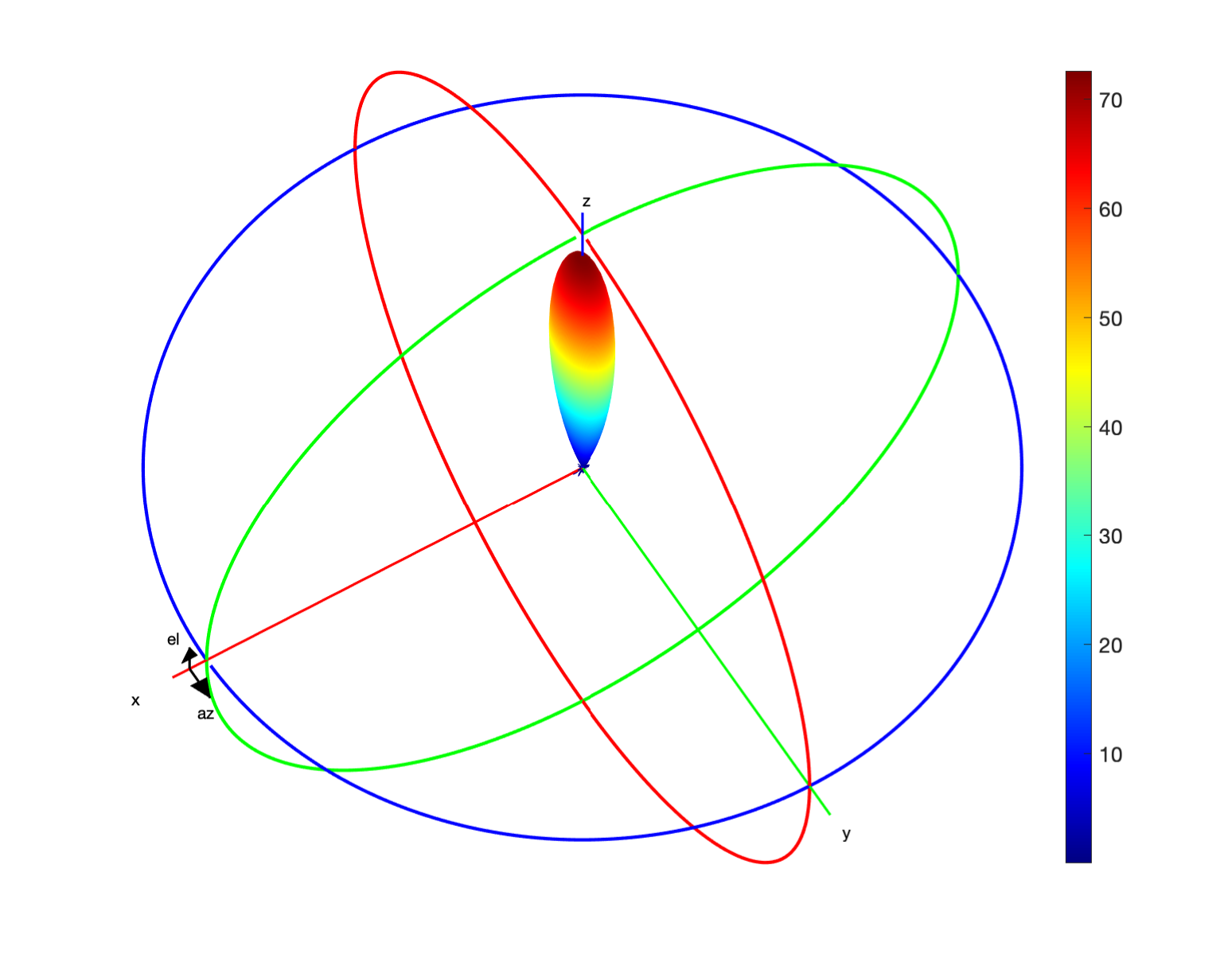}
\caption{3D pattern (linear scale) obtained with the DSA ($\phi=0^{\circ}$). Same configuration of Fig.~\ref{Fig:PatternDSA}.} 
\label{Fig:3DPattern}
\end{figure}

\begin{figure}[t]
\centering 
\centering\includegraphics[width=0.95\columnwidth]{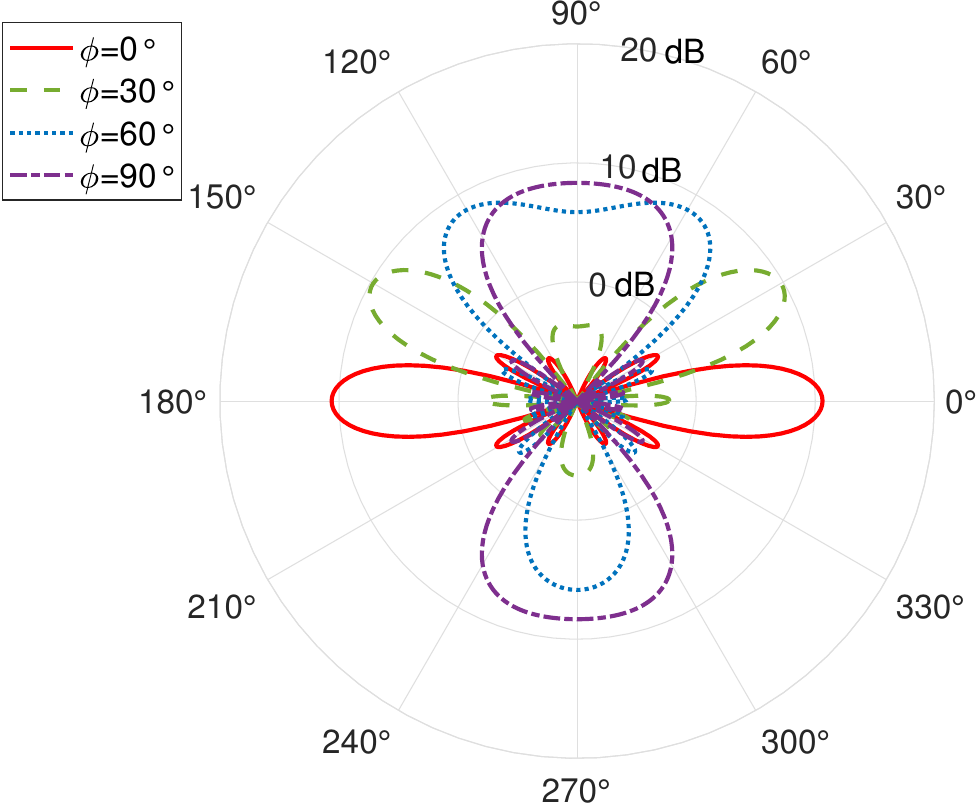}
\caption{Radiation pattern of a conventional ULA with $\Na=6$ active elements.} 
\label{Fig:PatternULA}
\end{figure}

\begin{figure}[t]
\centering 
\centering\includegraphics[width=0.9\columnwidth]{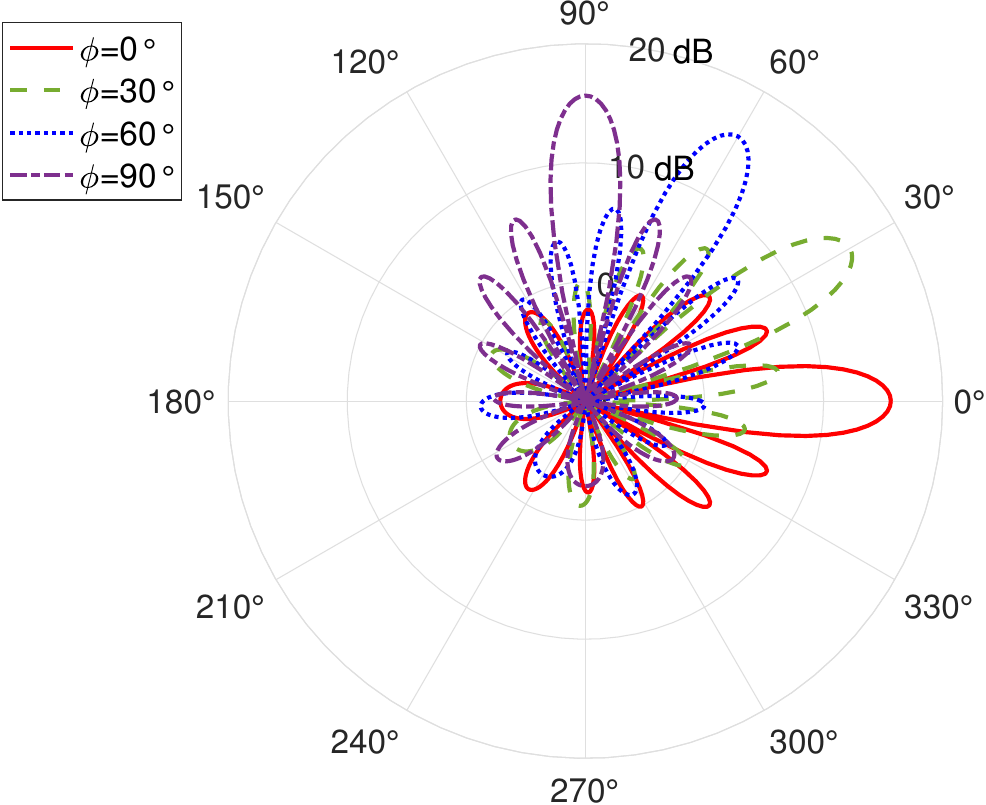}
\caption{Radiation pattern of a conventional UCA with $\Na=36$ active elements.} 
\label{Fig:PatternUCA}
\end{figure}

\subsection{Beam Forming and Superdirectivity}
We first investigate the beam forming capabilities of the \ac{DSA} using one RF chain ($\Na=1$ active antenna element) and the deployment of the scatterers according to $L$ concentric cylinders with incremental radius of step  $\Delta_L\,$ and scatterers separated of $\lambda/2$ within each cylinder along the circle and the $y$ axis (see Fig.~\ref{Fig:Example1}). We denote with $L_{\text{R}}$ the number of vertical rings composing each cylinder.
The cylindric structure appears very appealing especially for its use in base stations or access points thanks to its circular symmetry. When $L_{\text{R}}=1$, the cylinder degenerates into a disk.  Obviously, other structures can be considered as well depending on the specific application. 

The target end-to-end channel matrix $\boldHopt$ with $K=108$ test points deployed on the $x-z$ plane at distance $d=100\,$m (far field) according to the use case in Sec.~\ref{Sec:Example1} was considered for 4 different steering angles, $0^{\circ}$, $30^{\circ}$, $60^{\circ}$, and $90^{\circ}$, respectively. Only the optimization Step 2 was performed with $N_{\text{i}}=1500$ iterations and setting $\bWd=\sqrt{R}$ (in this case $\bWd$ is a scalar and no digital processing takes place).

In Fig.~\ref{Fig:PatternDSA}, the radiation diagrams for the 4 steering angles obtained with $\Delta_L=\lambda/4$, $L=5$, $L_{\text{R}}=1$ (disk shape), corresponding to $\Ns=121$ scatterers are shown.    
As it can be observed, the gain of the \ac{DSA} is independent of the angle and it is about $18.6\,$dB. In addition, limited back radiation is  obtained without the need to insert a ground plane that would impede the steering in the angle range $[90^{\circ} - 270^{\circ}]$. 
An example of the corresponding 3D radiation diagram for steering angle  $0^{\circ}$ is illustrated in Fig.~\ref{Fig:3DPattern}.   
 
As pointed out in the Introduction, the performance of any \ac{SIM} is constrained by the final layer surface of the structure, therefore it is upper bounded by that one can obtain by replacing the \ac{SIM} with a full digital planar or linear array with the same layout and number of elements.  
Therefore, for the sake of comparison, we consider the radiation diagrams obtained using a standard full-digital \ac{ULA} and a \ac{UCA}  with the same aperture of the \ac{DSA} ($3.2\,$cm diameter) in the $x-z$ plane. The corresponding radiation diagrams are reported in Figs.~\ref{Fig:PatternULA} and \ref{Fig:PatternUCA}, respectively. These arrays require $\Na=6$ and $\Na=36$ active antennas compared to only one active antenna of the \ac{DSA}.
As well-known, in \acp{ULA}  the gain degrades when approaching $90^{\circ}$ from a maximum value of $10.6\,$dB (about $10\log_{10}(\Na \, G_{\text{d}})$, with $G_{\text{d}}=3/2$ the gain of the Hertzian dipole) and symmetric back radiation is present. Instead, with the \ac{UCA} the gain ($15.6\,$dB) is insensitive to the steering direction at the expense of higher side lobes. Nevertheless, in both cases the proposed \ac{DSA} exhibits superdirectivity capability with an additional gain of $8\,$dB and $3\,$dB, respectively. 
Contrarily to standard arrays, where superdirectivity is obtained only in the end-fire direction \cite{IvrNos:14}, here notably superdirectivity is equally obtained in all steering directions.


\begin{table}[tp]
\caption{Cylindric shape. $L=5$.}
\begin{center}
\begin{tabular}{|c|c|c|c|}
\hline
 Configuration & Gain (dB) & $\Ns$ & $Q$-factor \\
 \hline
$\Delta_L=\lambda/8$,  $L_{\text{R}}=1$& 15 & 89 & 1.1 \\
$\Delta_L=\lambda/6$, $L_{\text{R}}=1$ & 16.7 & 100 & 1.25 \\ 
$\Delta_L=\lambda/4$, $L_{\text{R}}=1$ & 18.6 & 121 & 0.91 \\
$\Delta_L=\lambda/2$, $L_{\text{R}}=1$ & 16 & 184 & 0.24 \\
$\Delta_L=\lambda$, $L_{\text{R}}=1$ & 15.7 & 310 & 0.24 \\
$\Delta_L=\lambda/4$, $L_{\text{R}}=3$ & 20.4 & 363 & 0.3 \\ 
\hline
\end{tabular}
\end{center}
\label{Table:Cylinder}
\end{table}

\begin{table}[tp]
\caption{Random shape. $L_{\text{R}}=1$}
\begin{center}
\begin{tabular}{|c|c|c|c|c|}
\hline
$\Ns$ & 60 & 121 & 242 & 484  \\ 
\hline
Gain (dB) & 12.3 & 17  & 18.8  & 18  \\
 $Q$-factor & 0.29 & 1  & 1.55  & 3  \\
 \hline
\end{tabular}
\end{center}
\label{Table:Random}
\end{table}

In Table~\ref{Table:Cylinder}, the impact of the cylinders spacing $\Delta_L\,$ and the number of rings $L_{\text{R}}$ is investigated. 
As it can be noticed, the maximum gain is obtained for  $\Delta_L=\lambda/4$ confirming  that superdirectivity can be achieved only by exploiting the mutual coupling between the antenna elements but, at the same time, that there is an optimum spacing value to be found. 
In fact, on one hand, higher spacing reduces the coupling and hence the contribution to radiation of those scatterers located far away from the active element. On the other hand, very close spacing tightens the coupling between the elements and reduces the \ac{DoF} available for optimization. 
A higher number of rings ($L_{\text{R}}=3$) further increases the directivity by thinning the radiation diagram in the $y$ direction. The corresponding $Q$-factor is in general relatively small denoting a wideband characteristic of the \ac{DSA} under the considered setting 
and there is no need to generate very high currents as happens in conventional superdirective end-fire arrays \cite{IvrNos:10}. 

An example using a random deployment of the scatterers within a disk as a function of their density is provided in Table~\ref{Table:Random} for a fixed diameter of $3.2\,$cm. As it can be seen, the achievable performance is similar to that of the regular cylindrical deployment when compared with the same number of scatterers even though in general higher values of $Q$-factor are obtained because of the possibility that couples of scatterers are very close to each other.
The analysis and optimization of scatterers deployment strategies is a topic that deserves future investigations.     

An interesting aspect worth consideration is the sensitivity of the performance to parameter mismatch that might arise in practical implementations. We evaluated the gain degradation for the configuration in Fig.~\ref{Fig:PatternDSA} when the actual implemented $n$th parameter of the \ac{DSA}, $\hat{\theta}_n$, is affected by an error $d_n$ modeled as a zero-mean  Gaussian random variable with standard deviation $\sigma_n$, i.e., $\hat{\theta}_n=\theta_n+d_n$. The corresponding gains obtained for  $\sigma_n=0$, $\sigma_n=1\%$ and $\sigma_n=10\%$ of the nominal absolute value $|\theta_n|$ are, respectively, $G=18.6\,$dB, $G=15.8\,$dB, and $G=14.5\,$dB, showcasing a $4\,$dB loss with a $10\%$ relative error.

\begin{figure}[!t]
\centering 
\centering\includegraphics[width=0.9\columnwidth]{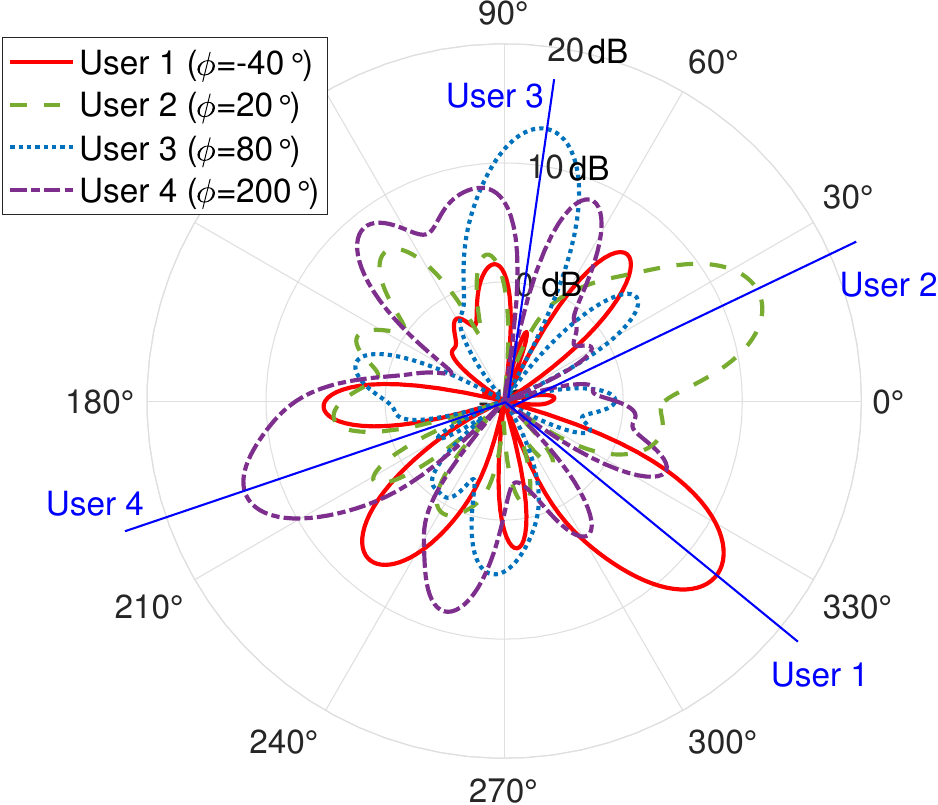}
\caption{Radiation diagram of a zero-forcing 4-user MISO system based on a DSA.} 
\label{Fig:PatternMISO}
\end{figure}

\subsection{Multi-user \ac{MISO}}
We consider now the scenario of use case in Sec.~\ref{Sec:Example2} illustrated in Fig.~\ref{Fig:Example2}, where a \ac{DSA} with $\Delta_L=\lambda/4$, $L_{\text{R}}=1$, $L=5$, $\Ns=121$, $\Na=4$ active antennas and $\Na$ RF chains is used to communicate simultaneously with $K=\Na$ single-antenna users located at a distance of $d=100\,$m and angles $-40^{\circ}$, $20^{\circ}$, $80^{\circ}$, and $200^{\circ}$ in the $x-z$ plane.  The \ac{DSA} has been optimized according to \eqref{eq:ZF} (zero forcing criterium) with $N_{\text{i}}=50$ and $N_{\text{alt}}=10$. 
The corresponding radiation diagrams are shown in Fig.~\ref{Fig:PatternMISO}. The coupling between users is impressive ($<-100\,$dB) at the expense of increased leakage in unwanted directions and a reduction of the gain of some dBs compared to the pattern in Fig.~\ref{Fig:PatternDSA} corresponding to a single user. 

Alternatively, one could use the approach in use case 1 of Sec.~\ref{Sec:Example1} to reduce the leakage in unwanted directions and increase the gain. However, our numerical investigations have revealed that the corresponding coupling between users would increase significantly. Therefore, the choice of the optimization method strongly depends on the specific application requirements in terms of gain versus coupling.

\subsection{Multi-layer MIMO}
Finally, we consider the scenario of use case 3 in Sec.~\ref{Sec:Example3} shown in Fig.~\ref{Fig:Example3}, where a multi-layer \ac{MIMO} link with a user located at $d=10\,$m equipped with a standard \ac{ULA} with $K=20$ elements spaced of $\lambda/2$ has to be established in \ac{NLOS} condition.
The \ac{DSA} is equipped with $\Na=4$ active antennas and $\Na$ RF chains to allow the transmission of up to 4 data streams (layers) according to channel characteristics.
The simulated \ac{NLOS} channel consists of 5 multi-paths caused by  the reflection of 5 scatterers located at angles $\phi=\{-43^{\circ},-14^{\circ},14^{\circ}, 43^{\circ}, 72^{\circ} \}$ and distance $5\,$m. 
The corresponding strongest singular values of the channel are $\bLambda=\diag{-35.6{\,\text{dB}},-42.4{\,\text{dB}},-43.7{\,\text{dB}},-46.3{\,\text{dB}}, -82{\,\text{dB}}, \ldots }\,$, which has clearly rank $r=4$.
The \ac{DSA} has the following parameters $\Delta_L=\lambda/4$, $L_{\text{R}}=1$, $\Ns=121$, and it is optimized according to the criterium in \eqref{eq:SVD} with $N_{\text{i}}=50$ and $N_{\text{alt}}=10$.

In Table~\ref{Table:Lambda}, the resulting coefficients $\hat{\bLambda}$ of the end-to-end channel is reported. The comparison between the diagonal and off-diagonal values indicates that the coupling between different layers is completely negligible, i.e., the channel is almost perfectly diagonalized. Compared to the singular values of the channel in $\bLambda$, the diagonal elements of $\hat{\bLambda}$ exhibit the same behavior with a loss of about $6\,$dB that can be ascribed to the fact that the antenna is not ideal.  
The effect of implementation errors is reported in the bottom part of the table for $1\%$ and $10\%$ relative error standard deviation, respectively. It can be seen that the effect of implementation errors becomes no longer negligible (coupling above $-10\,$dB) with $10\%$ relative error.   

\begin{table}[tp]
\caption{${\left [ \hat{\Lambda} \right ]  }_{n, k}$ in dB, for $n, k=1, 2, 3, 4$.}
\begin{center}
\begin{tabular}{|c|c|c|c|}
\hline
\multicolumn{4}{|c|}{Perfect implementation }    \\
\hline
-41.6  & -177  & -190 & -175 \\
 -170 & -48 & -185 & -170\\
 -183 & -184 &  -49.7 & -182 \\
 -164 & -166 & -180  & -52.3 \\
 \hline
 \hline
\multicolumn{4}{|c|}{$1\%$ relative implementation error}    \\
\hline
-41.9 &  -90.6  & -80  &  -80.5 \\
-89  & -48.7  & -98.5  & -86.7 \\ 
-87  & -111  & -50  & -84.9 \\
-120 & -101  & -89.7  & -52.5  \\
\hline
 \hline
\multicolumn{4}{|c|}{$10\%$ relative implementation error}  \\
\hline
-47  & -56.7  & -59.4  & -62.2 \\
  -66.6  & -48.3  & -61.6  & -65.3 \\
  -67  & -61.4  & -52.4  & -64.9 \\
  -57.2  & -74.7  & -62.8  & -55.3 \\
\hline
\end{tabular}
\end{center}
\label{Table:Lambda}
\end{table}


\section{Conclusion}
\label{Sec:Conclusion}

In this paper, we have introduced the concept of the \acf{DSA} as an appealing technology aimed at shifting part of the signal processing from the digital domain to the \ac{EM} domain. This is achieved through the joint optimization of \ac{EM} processing and radiation of active and reconfigurable scattering elements interacting in the reactive near field.
We have shown that the \ac{DSA} provides unprecedented flexibility in managing the \ac{EM} field by reducing or circumventing digital processing at the baseband, thereby surpassing the flexibility offered by classical \ac{MIMO} and recent \ac{SIM} implementations. Consequently, the reduction in \ac{RF} chains and digital processing facilitates the realization of energy-efficient, low-latency, and cost-effective holographic \ac{MIMO} systems, then addressing the sustainability concerns of future wireless networks.

Although numerous implementation challenges must be addressed before \acp{DSA} can become a reality, such as the design of flexible reconfigurable scatterers, we hope that our promising theoretical results will inspire further research into technologies for the efficient implementation of \acp{DSA} as a key enabler of holographic \ac{MIMO} within the \ac{ESIT} paradigm.

\section*{Acknowledgment}
This work was supported by the European Union under the Italian National Recovery and Resilience Plan (NRRP) of NextGeneration EU, partnership on ``Telecommunications of the Future" (PE00000001 - program ``RESTART"), and by the HORIZON-JU-SNS-2022-STREAM-B-01-03 6G-SHINE project (Grant Agreement No. 101095738).

\ifCLASSOPTIONcaptionsoff
\fi
\bibliographystyle{IEEEtran}
\bibliography{IEEEabrv,BiblioDD,MetaSurfaces,EMInformationTheory,IntelligentSurfaces,MassiveMIMO,MIMO,EMTheory,WINS-Books,Vari}

\end{document}

%% file: acronyms.tex
\acrodef{SIM}{stacked intelligent metasurface}

\acrodef{DAC}{digital-to-analog converter}

\acrodef{DMA}{dynamic metasurface antenna}

\acrodef{SINR}{signal-to-interference noise ratio}

\acrodef{ESIT}{electromagnetic signal and information theory} 

\acrodef{ELAA}{extremely large antenna arrays} 

\acrodef{DSA}{dynamic scattering array}

\acrodef{ULA}{uniform linear array}

\acrodef{UCA}{uniform circolar array}

\acrodef{IIoT}{industrial Internet-of-things}

\acrodef{IT}{information theory}

\acrodef{SRE}{smart radio environment}

\acrodef{EMO}{electromagnetic object}

\acrodef{SVD}{singular value decomposition}

\acrodef{PSWF}{prolate spheroidal wave function}

\acrodef{CR}{channel response}

\acrodef{BS}{base station}

\acrodef{MS}{mobile station}

\acrodef{UE}{user equipment}

\acrodef{MIMO}{multiple-input multiple-output}

\acrodef{MISO}{multiple-input single-output}

\acrodef{RIS}{reconfigurable intelligent surface}

\acrodef{IRS}{intelligent reconfigurable surface}

\acrodef{LIS}{large intelligent surface}

\acrodef{MIS}{medium intelligent surface}

\acrodef{SIS}{small intelligent surface}

\acrodef{DoF}{degrees-of-freedom}

\acrodef{AF}{amplify \& forward}

\acrodef{DF}{detect \& forward}

\acrodef{JF}{just forward}

\acrodef{CSI}{channel state information}

\acrodef{RV}{random variable}

\acrodef{i.i.d.}{independent, identically distributed}

\acrodef{PSD}{power spectral density}

\acrodef{PDF}{probability distribution function}

\acrodef{CDF}{cumulative distribution function}

\acrodef{ch.f.}{characteristic function}

\acrodef{AWGN}{additive white Gaussian noise}

\acrodef{RSSI}{received signal strength indicator}

\acrodef{SNR}{signal-to-noise ratio}

\acrodef{LRT}{likelihood ratio test}

\acrodef{GLRT}{generalized likelihood ratio test}

\acrodef{GML}{generalized maximum likelihood}

\acrodef{LOS}{line-of-sight}

\acrodef{NLOS}{non-line-of-sight}

\acrodef{GDOP}{geometric dilution of precision}

\acrodef{GPS}{Global Positioning System}

\acrodef{FIM}{Fisher information matrix}

\acrodef{PEB}{position error bound}

\acrodef{WSN}{Wireless Sensor Network}

\acrodef{MAC}{medium access control}

\acrodef{RSS}{received signal strength}

\acrodef{RTT}{round-trip time}

\acrodef{MIMO}{multiple-input multiple-output}

\acrodef{MF}{matched filter}

\acrodef{ED}{energy detector}

\acrodef{ML}{maximum likelihood}

\acrodef{NL}{nonlinear}

\acrodef{MSE}{mean square error}

\acrodef{RMSE}{root mean square error}

\acrodef{ppm}{part-per-million}

\acrodef{PRP}{pulse repetition period}

\acrodef{ACK}{acknowledge}

\acrodef{UWB}{ultrawide bandwidth}

\acrodef{TNR}{threshold-to-noise ratio}

\acrodef{LOS}{line-of-sight}

\acrodef{LS}{least squares}

\acrodef{IR-UWB}{impulse radio UWB}

\acrodef{FCC}{Federal Communications Commission}

\acrodef{TH}{time-hopping}

\acrodef{PPM}{pulse position modulation}

\acrodef{PAM}{pulse amplitude modulation}

\acrodef{MUI}{multi-user interference}

\acrodef{PDP}{power delay profile}

\acrodef{PPP}{Poisson point process}

\acrodef{DS}{delay spread}

\acrodef{CED}{channel excess delay}

\acrodef{BPZF}{band-pass zonal filter}

\acrodef{SIR}{signal-to-interference ratio}

\acrodef{RFID}{radio frequency identification}

\acrodef{WPAN}{wireless personal area networks}

\acrodef{WWLB}{Weiss-Weinstein lower bound}

\acrodef{DP}{direct path}

\acrodef{MF}{matched filter}

\acrodef{MMSE}{minimum-mean-square-error}

\acrodef{SBS}{serial backward search}

\acrodef{NBI}{narrowband interference}

\acrodef{WBI}{wideband interference}

\acrodef{INR}{interference-to-noise ratio}

\acrodef{CIR}{channel impulse response}

\acrodef{ISI}{inter-symbol interference}

\acrodef{CPR}{channel pulse response}

\acrodef{LRT}{likelihood ratio test}

\acrodef{MUI}{multi-user interference}

\acrodef{EM}{electromagnetic}

\acrodef{CW}{continuous wave}

\acrodef{RF}{radiofrequency}

\acrodef{FCC}{Federal Communications Commission}

\acrodef{EIRP}{effective radiated isotropic power}

\acrodef{RCS}{radar cross section}

\acrodef{BAV}{balanced antipodal Vivaldi}

\acrodef{PRake}{partial Rake}

\acrodef{RTLS}{real time locating system}

\acrodef{CRB}{Cram\'{e}r-Rao bound}

\acrodef{ZZB}{Ziv-Zakai bound}

\acrodef{TOA}{time-of-arrival}

\acrodef{TOF}{time-of-flight}

\acrodef{WSN}{wireless sensor network}

\acrodef{MAC}{medium access control}

\acrodef{RSS}{received signal strength}

\acrodef{TDOA}{time difference-of-arrival}

\acrodef{RF}{radiofrequency}

\acrodef{RTT}{round-trip time}

\acrodef{AOA}{angle-of-arrival}

\acrodef{MF}{matched filter}

\acrodef{ED}{energy detector}

\acrodef{ML}{maximum likelihood}

\acrodef{MUR}{Multistatic radar}

\acrodef{HDSA}{high-definition situation-aware}

\acrodef{RRC}{root raised cosine}

\acrodef{OFDM}{orthogonal frequency division multiplexing}

\acrodef{IF}{intermediate frequency}

\acrodef{PHY}{physical layer}

\acrodef{S-V}{Saleh-Valenzuela}

\acrodef{UHF}{ultra-high frequency}

\acrodef{PR}{pseudo-random}

\acrodef{SoC}{System on Chip}

\acrodef{SoP}{System on Package}

\acrodef{SPMF}{Single-Path Matched Filter}

\acrodef{IMF}{Ideal Matched Filter}

\acrodef{SCR}{signal-to-clutter ratio}

\acrodef{BEP}{bit error probability}

\acrodef{BER}{bit error rate}

\acrodef{WSR}{wireless sensor radar}

\acrodef{HPBW}{half power beam width}

\acrodef{LEO}{localization error outage}

\acrodef{WSS}{wide-sense stationary}

\acrodef{TR}{time-reversal}

\acrodef{WSSUS}{WSS with uncorrelated scattering}

\acrodef{GP}{Gaussian process}

\acrodef{IMU}{inertial measurement unit}


%% file: def_EM_SP.tex
\newcommand{\Real}[1]{\Re \left \{ #1\right \}}
\newcommand{\Imag}[1]{\Im \left \{ #1\right \}}

\newcommand{\EX}[1] {{\mathbb{E}}\left\{{#1}\right\}}

\newcommand{\boldA} {{\bf{A}}}

\newcommand{\bolds} {{\bf{s}}}

\newcommand{\boldp} {{\bf{p}}}

\newcommand{\boldV} {{\bf{V}}}

\newcommand{\boldt} {{\bf{t}}}
\newcommand{\boldU} {{\bf{U}}}

\newcommand{\boldH} {{\bf{H}}}

\newcommand{\boldM} {{\bf{M}}}

\newcommand{\boldC} {{\bf{C}}}

\newcommand{\boldI} {{\bf{I}}}
\newcommand{\boldr} {{\bf{r}}}

\newcommand{\boldx} {{\bf{x}}}
\newcommand{\boldy} {{\bf{y}}}

\newcommand{\diag}[1]{{\rm diag} \left ( #1 \right )}

\newcommand{\argmin}[1]{\underset{{#1}}{\operatorname{arg \, min}}}
\newcommand{\minimize}[1]{\underset{{#1}}{\operatorname{minimize}}}

\newcommand{\ctranspose}{^{\mathsf{H}}}
\newcommand{\transpose}{^{\mathsf{T}}}

\newcommand{\versorp} {\hat{\bf{p}}}

\newcommand{\versorx} {{\hat{\bf{x}}}}
\newcommand{\versory} {{\hat{\bf{y}}}}
\newcommand{\versorz} {{\hat{\bf{z}}}}

\newcommand{\Na} {N_{\text{A}}}
\newcommand{\Ns} {N_{\text{S}}}

\newcommand{\Ptx} {P_{\text{T}}}
\newcommand{\Prx} {P_{\text{R}}}
\newcommand{\Prad} {P_{\text{rad}}}
\newcommand{\Preact} {P_{\text{react}}}

\newcommand{\Gr} {G_{\text{R}}}

\newcommand{\bi} {\mathbf{i}}
\newcommand{\bit} {\bi_{\text{T}}}
\newcommand{\bia} {\bi_{\text{A}}}
\newcommand{\bis} {\bi_{\text{S}}}

\newcommand{\bv} {\mathbf{v}}
\newcommand{\bvt} {\bv_{\text{T}}}
\newcommand{\bva} {\bv_{\text{A}}}

\newcommand{\bvs} {\bv_{\text{S}}}

\newcommand{\bOmega} {\boldsymbol{\Omega}}
\newcommand{\bOmegak} {\boldsymbol{\Theta}_k}
\newcommand{\btheta} {\boldsymbol{\theta}}

\newcommand{\bLambda} {\boldsymbol{\Lambda}}

\newcommand{\bZ} {\mathbf{Z}}
\newcommand{\bZa} {\mathbf{Z}_{\text{A}}(\btheta)}
\newcommand{\bZaa} {\mathbf{Z}_{\text{AA}}}
\newcommand{\bZas} {\mathbf{Z}_{\text{AS}}}
\newcommand{\bZsa} {\mathbf{Z}_{\text{SA}}}
\newcommand{\bZss} {\mathbf{Z}_{\text{SS}}}
\newcommand{\bZs} {\mathbf{Z}_{\text{S}}(\btheta)}
\newcommand{\Zs}[1] {Z_{\text{S}_{#1}} \left (\theta_{#1} \right )}

\newcommand{\bZt} {\mathbf{Z}_{\text{T}}}

\newcommand{\bWd} {\mathbf{W}_{\text{D}}}

\newcommand{\bWdhat} {{\hat{\mathbf{W}}}_{\text{D}}}

\newcommand{\bWem} {\mathbf{W}_{\text{EM}}(\btheta)}
\newcommand{\bWemhat} {\mathbf{W}_{\text{EM}}(\hat{\btheta})}

\newcommand{\bHc} {\mathbf{H}_{\text{c}}}
\newcommand{\bhc} {{\mathbf{h}_{\text{c}}}}
\newcommand{\Hc} {{H_{\text{c}}}}

\newcommand{\bZm} {\mathbf{Z}_{\text{M}}(\btheta)}

\newcommand{\Gej}[1] {\mathbf{G}_{{EJ}}\left ( {#1} \right )}

\newcommand{\EMsymb} {\mathsf}

\newcommand{\Em} {{\EMsymb E}}
\newcommand{\Jm} {{\EMsymb J}}

\newcommand{\boldzero} {\boldsymbol{0}}

\newcommand{\kappazero} {\kappa_0}

\newcommand{\scalprod} {\boldsymbol{\cdot}}

\newcommand{\boldHopt} {{\bf{H}_{\text{opt}}}}